\documentstyle[preprint,aps,epsfig]{revtex}  

\begin{document}
\draft
\preprint{\today}


\title{Proton-neutron interactions in
 $N \approx Z$ nuclei}
\vspace{10 mm}

\author{
 K. Kaneko$^{1}$ and M. Hasegawa$^{2}$}
\address{
$^{1}$Department of Physics, Kyushu Sangyo University,
Matsukadai, Fukuoka 813-8503 Japan\\
$^{2}$Laboratory of Physics, Fukuoka Dental College, Fukuoka 
814-0193 Japan}

\maketitle

\begin{abstract}

   Proton-neutron ($p-n$) interactions and their various aspects in
$N\approx Z$ nuclei of $g_{9/2}$- and $f_{7/2}$ subshell are studied
using a schematic model interaction with four force parameters proposed
recently.
It is shown that the model interaction well reproduces observed 
physical quantities: the double differences of binding energies, 
symmetry energy, Wigner energy, odd-even mass 
difference and separation energy, which testifies to the reliability of the model interaction
and its $p-n$ interactions. 
 First of all, the double differences of binding energies are used
for probing the p-n interactions.  The analysis reveals
different contributions of the isoscalar and isovector $p-n$
pairing interactions to two types of double differences of binding
energies, and also indicates the importance of a unique form of
isoscalar $p-n$ pairing force with all $J$ components.
Next, it is shown that this p-n force is closely related to
the symmetry energy and the Wigner  energy.  
Other calculations demonstrate the significant roles of $p-n$ interactions
in the odd-even mass difference and in the separation energy at $N=Z$.

\vspace*{10mm}
PACS number(s): 21.10.Dr;21.10.Hw;21.60.-n;21.60.Cs
\end{abstract}



\newpage

\section{Introduction}
With the advent of radioactive nuclear beams, the properties of nuclei beyond 
the proton stability line have attracted experimental and theoretical attention
 in recent years. Special interest is devoted to a unique aspect 
 originating from the fact 
 that protons and neutrons occupy the same orbits in nuclei with 
 $N \approx Z$( see \cite{Goodman} for review ). Consequently, one expects a strong 
 proton-neutron ($p-n$) interaction because of the large spatial overlaps 
 between proton and neutron single-particle wave functions. 
The correlation energies related to the $p-n$ interaction have been extracted from the experimental binding energies \cite{Janecke1,Jensen,Janecke2,Dussel}. 
A double difference of binding energies has been analyzed with the aim of 
providing input to semiempirical mass formulas \cite{Janecke1,Jensen,Janecke2},
 and with relation to the clustering of nucleons as elementary modes of 
 excitation in nuclei \cite{Dussel}. It has recently been used for study of  
 the p-n interactions \cite{Zhang,Zaochun}, and discussed in terms of 
 schematic and realistic shell model calculations \cite{Brenner}. 
 This approach using the double difference of binding energies may provide 
 details of the $p-n$ interactions, the isoscalar ($\tau=0$) and isovector 
 ($\tau=1$) $p-n$ interactions. 
The $p-n$ part of the isovector pairing correlation near $N=Z$ 
has been studied 
in terms of algebraic model \cite{Hecht} and compared with shell model Monte 
Carlo calculations \cite{Engel1,Langanke}. 
On the other hand, there are many discussions of the roles of the isoscalar 
pairing interactions (see Refs.\cite{Satula1,Satula2,Engel2,Rudolph,Dean,Kaneko1} for instance). 

The experimental data indicate \cite{Janecke1,Jensen,Janecke2} that the 
symmetry energy accompanied by the so-called Wigner energy behaves according to the $T(T+1)$ dependence ($T=|Tz|$). This form could come from the 
isospin-invariant Hamiltonian. It has been recently proposed that the Wigner 
energy originates in the isoscalar pairing interaction 
\cite{Satula1,Satula2}. They pointed out that the Wigner energy cannot be 
solely explained in terms of correlations between the $J=1$ isoscalar $p-n$ 
pairs, and the isoscalar $p-n$ pairs with the other $J$ 
contribute significantly 
\cite{Satula2}. In fact, a recent shell model calculation \cite{Poves} with 
the $J=0$ isovector and $J=1$ isoscalar pairing forces in the $N \approx Z$ 
$fp$ shell nuclei cannot explain the magnitude of the Wigner energy. On the 
other hand, another paper \cite{Vogel1} has discussed the fact that the
 degeneracy of the $T=0$ and $T=1$ states in odd-odd nuclei with $N=Z$
 is produced by a balance of the symmetry energy and the $J=0$
 isovector pairing correlation.
 The lowering of the $T=0$ states in $N=Z$ odd-odd nuclei, according to
 our investigation \cite{Kaneko2,Hasegawa}, is caused by a unique form
 of isoscalar ($p-n$) pairing force including all $J$ components.
 This result is consistent with that of Satula et al. \cite{Satula1,Satula2}
 The uniqueisoscalar $p-n$ pairing force, which can be 
 expressed in a simple form
 including the $T(T+1)$ term, manifests a close relation to the
 symmetry energy.  We shall discuss this matter by a concrete
 calculation in this paper.

The odd-even mass difference (OEMD), the extra binding energy of a nucleus 
relative to its neighbors, is known to be an obvious experimental evidence 
of the  pairing correlation \cite{Bohr1}. The pairing phenomena are well 
understood in terms of the proton-proton ($p-p$) or 
neutron-neutron ($n-n$) pair 
condensate, and described by the Bardeen-Cooper-Schriefer(BCS) theory 
\cite{Bardeen}. The OEMD is often interpreted as a measure of the pairing gap 
(following the relation $12/A^{1/2}$ on the average) in medium-heavy and heavy 
nuclei. The OEMD displays, however a different feature in $N \approx Z$ nuclei 
such as a special increase at $N=Z$. 
An ordinary estimation of the neutron or proton pairing gap from the OEMD is
 not applicable to these nuclei. On the other hand, it has recently been 
 discussed that the OEMD in light nuclei is affected by deformation as well as 
 $J=0$ pairing correlation \cite{Mann,Hakkinen,Satula3}. 
 A further investigation of the OEMD should be made in $N \approx Z$
 nuclei.  We shall discuss the influence of the $p-n$ interactions
 on the OEMD.

The development of recent radioactive nuclear beams facilities provides 
unstable nuclei beyond the line of proton stability. Experimental and 
theoretical investigations of proton emitters are increasing. Such phenomena 
allow a test of the various models on the proton-rich side. For $N \approx Z$ 
nuclei, one can expect that the $p-n$ interactions also influence the 
separation energy. Since the $p-n$ force is considered to be attractive, it 
might increase the separation energy. In fact, the calculated separation 
energies by all models without the $p-n$ interaction are smaller than those of 
experiments at the $N=Z$ nuclei. 

We need a reliable effective interaction to study the nuclear properties 
mentioned above. We have proposed an extension of the $P+QQ$ model with four 
forces \cite{Hasegawa}, which reproduces quite well the experimental binding 
energies and energy spectra in $N \approx Z$ nuclei of $g_{9/2}$ and 
$f_{7/2}$ subshells. 
This model interaction including different types of
 $p-n$ forces is very suitable for our purpose to study various aspects
 of the $p-n$ interactions.
The main purpose of this paper is to study 
the $p-n$ interactions, analyzing the 
double differences of binding energies, and to check the validity of our model 
examining various quantities such as the symmetry energy, the Wigner energy, 
the odd-even mass difference and the separation energy in nuclei near $N=Z$. 

  The paper is organized as follows. In Sec. II, we first review
 our model proposed in the previous paper. Sec. III contains the
 analysis of the double differences of binding energies to probe
 the $p-n$ interactions. The symmetry energy and the Wigner energy are
 discussed in Sec. IV. In Sec. V, the odd-even mass differences
 are analyzed in detail, and the two-proton separation energies are
 calculated in Sec. VI. Finally, Sec. VII gives conclusions.

\section{
 Model interaction }

 We have proposed the following effective interaction extended from
 the $P+QQ$ force which is composed of four isospin-invariant forces
 (see Ref. \cite{Hasegawa} in detail):
 \begin{eqnarray}
  & & H = H_{\rm sp} + V_{\rm int}  \label{eq:6} \\
  & {} & H_{\rm sp} = \sum_{\alpha \rho} \epsilon_a c^\dagger_{\alpha 
\rho}
  c_{\alpha \rho}, \label{eq:7} \\
 & {} & V_{\rm int} =  V(P_0) + V(QQ) + V(P_2) 
  + V^{\tau=0}_{\pi \nu}, \label{eq:8} \\
 & {} & V(P_J)= -{1 \over 2} g_J \sum_{M \kappa}
  \sum_{a \leq b} P^\dagger_{JM1\kappa}(ab)
  \sum_{c \leq d} P_{JM1\kappa}(cd),   \label{eq:9} \\
 & {} & V(QQ)=-{1 \over 2} \chi : \sum_\mu \sum_{ab \rho}
  Q^\dagger_{2\mu \rho}(ab) \sum_{cd \rho^\prime}
  Q_{2\mu \rho^\prime}(cd):,  \label{eq:10} \\
 & {} & V^{\tau =0}_{\pi \nu}=-k^0 \sum_{a \leq b}
   \sum_{JM} A^\dagger_{JM00}(ab) A_{JM00}(ab),  \label{eq:11}
\end{eqnarray}
with
\begin{eqnarray}
 P^\dagger_{JM1\kappa}(ab) & = & p_J(ab) A^\dagger_{JM1\kappa}(ab),
   \label{eq:12} \\
 Q^\dagger_{2\mu \rho}(ab) & = & q(ab) B^\dagger_{2\mu \rho}(ab),
   \label{eq:13} \\
 A^\dagger_{JM \tau \kappa}(ab) & = & \sum_{m_\alpha m_\beta}
 \langle j_a m_\alpha j_b m_\beta |JM \rangle \sum_{\rho \rho^\prime}
 \langle \frac{1}{2} \rho \frac{1}{2} \rho^\prime |\tau \kappa \rangle
 { c^\dagger_{\alpha \rho} c^\dagger_{\beta \rho^\prime} \over
  \sqrt{1 + \delta_{ab} } },   \label{eq:14} \\
 B^\dagger_{2\mu \rho}(ab) & = & \sum_{m_\alpha m_\beta}
 \langle j_a m_\alpha j_b {\overline m_\beta} |2\mu \rangle
 c^\dagger_{\alpha \rho} (-)^{j_b - m_\beta}
 c_{\beta \rho} , \label{eq:15}
\end{eqnarray}
 where $p_0(ab)$=$\sqrt{ (2j_a+1) } \delta_{ab}$
 and $p_2(ab)$=$q(ab)$=$(a\|r^2Y_2\|b)/\sqrt{5}$.
 We use the notation $JM$ and $\tau \kappa$ for the spin and isospin
 of a nucleon pair, respectively.
 The subscript $\rho$ denotes the $z$ components of isospin
 $\pm \frac{1}{2}$.  We also use the notation $\rho=\pi$ for a proton
 and $\rho=\nu$ for a neutron.

  Here $H_{\rm sp}$ is a single-particle Hamiltonian and $V_{\rm int}$
 contains the four forces: $V(P_0)$ stands for the isovector monopole
 pairing force, $V(QQ)$ for the isoscalar quadrupole-quadrupole force,
 $V(P_2)$ for the isovector quadrupole pairing force and 
 $V^{\tau =0}_{\pi \nu}$ for the $J$-independent isoscalar $p-n$ force.
 The first two forces in the interaction (\ref{eq:8}) are an extension
 of the conventional $P+QQ$ force to the isospin-invariant one.
 The p-n part of the monopole and quadrupole pairing forces as well as
 the $p-n$ component of the quadrupole-quadrupole force would be important for
 $N \approx Z$ nuclei.  The last $p-n$ force is very important for reproducing
 the experimental binding energy. 
 It is important to note that $V^{\tau =0}_{\pi \nu}$ can be expressed as
 a simple form
\begin{eqnarray}
 & {} & V^{\tau =0}_{\pi \nu}= - \frac{1}{2}k^0 \{ {{\hat n} \over 2}
     ({{\hat n} \over 2}+1) - {\hat {T^2}} \},
    \label{eq:16}
\end{eqnarray}
 where ${\hat n}$ denotes the total number operator of
 valence nucleons (${\hat n}={\hat n}_{p}+{\hat n}_{n}$) and
 ${\hat T}$ is the total isospin operator. 
Our $p-n$ interaction is composed of
 four different components and hence is useful in analyzing their
 respective contributions to various physical quantities.
 
   We applied the above hamiltonian to examine the binding energies and
 energy spectra of nuclei with $A=82-100$ and $A=42-50$ \cite{Hasegawa}.
 We adopted only the $g_{9/2}$ shell for nuclei with $A$=80-100
 regarding the $Z=N=40$ core as inactive, and the $f_{7/2}$ shell
 for nuclei with $A$=40-50 regarding the $Z=N=20$ core as inactive. 
 It may be necessary to extend these model spaces for quantitative 
 discussion.  Our calculation
 itself indicates the insufficiency of the model space $(f_{7/2})^n$
 about energy spectra.  We used an extended model space
 $(p_{1/2},g_{9/2})^n$ when comparing calculated energy levels with
 observed ones in nuclei with $A \approx 90$, while experimental data near
 $A=80$ seem to demand a further extension of the model
 space.  The previous paper, however, has shown that the single $j$
 shell model is bearable for semiquantitative discussion about the
 nuclear binding energy.  This simple model makes it possible to 
 clearly see the roles of respective $p-n$ interactions.
 We therefore employ the same single $j$ shell model as that used
 in Ref. \cite{Hasegawa}, where the following force strengths are used:
 \begin{eqnarray}
  & &g_{0}=0.26, \quad X=\chi \{ q(g_{9/2} g_{9/2}) \}^{2}=
1.50,
   \nonumber \\
  & & G_{2}=\frac{1}{2}g_{2} \{ q(g_{9/2} g_{9/2}) \}^{2}=
0.35,
   \quad k^{0} =0.925  \quad \mbox{in MeV for the $g_{9/2}$ shell 
region},
    \label {eq:100}
 \end{eqnarray}
 \begin{eqnarray}
  & &g_{0}=0.59, \quad X=\chi \{ q(f_{7/2} f_{7/2}) \}^{2}=
1.20,
   \nonumber \\
  & & G_{2}=\frac{1}{2}g_{2} \{ q(f_{7/2} f_{7/2}) \}^{2}=
0.90,
   \quad k^{0} =1.90  \quad \mbox{in MeV for the $f_{7/2}$ shell 
region}.
    \label{eq:110}
 \end{eqnarray}
 Our model with these sets of parameters is considered to be reliable
 for studying the p-n interactions in connection with physical quantities
 related to the binding energy in the $g_{9/2}$  and $f_{7/
2}$
 shell nuclei. 
 The mass dependence of the force parameter $k^{0}$ is taken into account
 in some cases, but it does not change qualitatively the result.

\section{Double differences of binding energies and p-n interactions}

We define the $m$th double difference of binding energies as follows:
\begin{eqnarray}
\delta V^{(m)}(Z,N)=\delta^{(m)}B(Z,N), \label{eq:101}
\end{eqnarray}
where $B(Z,N)$ is the nuclear binding energy. Here the operator $\delta^{(m)}$ is defined as
\begin{eqnarray}
\delta^{(m)}f(Z,N)=-\frac{1}{m^{2}}[f(Z,N)-f(Z,N-m)  \nonumber  \\
-f(Z-m,N)+f(Z-m,N-m)]. \label{eq:102}
\end{eqnarray}
The double difference of binding energies, $\delta V^{(1)}(Z,N)$, was 
introduced for investigating the semiempirical mass formula 
\cite{Janecke1,Jensen,Janecke2}. This quantity is expected to roughly 
represent the $p-n$ interactions between the last proton and 
neutron from the form of Eq. (\ref{eq:102}). Figure 1(a) shows 
the plot of $\delta V^{(1)}(Z,N)$ 
as a function of $A=N+Z$ for nuclei in the mass region $A$=16-165. 
Experimental data are taken from Ref.\cite{Table}. We see two separate groups 
in Fig. 1(a), namely, one is for the even-$A$ nuclei (dots) and the other is 
for the odd-$A$ nuclei (crosses).
In both cases, shell effects at $Z$ or $N=28,40,50,82$ are present, while the 
patterns of dots and crosses are symmetric with respect to the average curve. 
It is now convenient to divide $\delta V^{(1)}(Z,N)$ into two parts: 
the average 
part of the even-$A$ and odd-$A$ nuclei and the deviation from it. As seen in 
Fig. 1(a), the former is approximately written as $I_{0}=40/A$ and the latter 
has opposite signs for the even-$A$ and odd-$A$ nuclei as follows:
\begin{eqnarray}
\delta V^{(1)}(Z,N) \approx I_{0} + (-1)^{A}I_{1}. \label{eq:103}
\end{eqnarray}
This expression was originally given by de-Shalit 
\cite{Janecke2,deShalit} in the earliest investigations of the effective p-n 
interactions. Equation (\ref{eq:103}) describes the staggering with respect to 
the isotopes for even-$A$ and odd-$A$ nuclei. Large values of 
$\delta V^{(1)}(Z,N)$ for even-$A$ nuclei (dots) near $N=Z$ below 
$A=80$ are notable.

The data of $\delta V^{(2)}(Z,N)$ are plotted in Fig. 1(b) as a function of 
$A=N+Z$. (Our definition of $\delta V^{(2)}$ has a sign 
opposite to that of Brenner's 
$et al.$ \cite{Brenner}.)
The values of $\delta V^{(2)}$ show a different behavior from $\delta V^{(1)}$. 
It is interesting that the staggering of $\delta V^{(1)}$ disappears in 
$\delta V^{(2)}(Z,N)$. 
We see large scatters of dots and crosses for $A<80$. These correspond to 
$\delta V^{(2)}$ of nuclei in $N \approx Z$, and the values of 
$\delta V^{(2)}$ at $N=Z$ are especially large. 
With decreasing mass $A$, $\delta V^{(2)}$ at $N=Z$ increases. If one neglects 
the dots and crosses in $N \approx Z$ nuclei, $\delta V^{(2)}$ varies rather 
smoothly. This smooth trend is clear for $A > 80$ and continues up to heavy 
nuclei.  This is due to the fact that there is no stable $N \approx Z$ nuclei 
with $A>80$. Figure 1(b) clearly indicates the smooth systematic decrease of 
$\delta V^{(2)}$ with increasing mass $A$, which can be traced by the curve 
$40/A$. 
The deviations from the curve $40/A$ are small, and shell structure is not 
found.
This general trend of $\delta V^{(2)}$ has long been known, and was discussed 
in several papers.  In a recent paper \cite{Brenner}, the dramatic spikes of 
$\delta V^{(2)}$ at $N=Z$ light nuclei were discussed in terms of both 
schematic and realistic shell model calculations, and the importance of 
$\tau=0$ $p-n$ interaction for the spikes was pointed out.
The following relationship is derived from Eqs. (\ref{eq:101}) and 
(\ref{eq:102}): 
\begin{eqnarray}
\delta V^{(2)}(Z,N)=\frac{1}{4}[\delta V^{(1)}(Z,N)+\delta V^{(1)}(Z,N-1) 
\nonumber \\
+\delta V^{(1)}(Z-1,N)+\delta V^{(1)}(Z-1,N-1)]. \label{eq:4}
\end{eqnarray}
Substituting the empirical relationship (\ref{eq:103}) 
into Eq. (\ref{eq:4}), 
in the large $A$ limit we get
\begin{eqnarray}
\delta V^{(2)}(Z,N) \approx I_{0}=40/A. \label{eq:5}
\end{eqnarray}
Thus, the systematic behavior of $\delta V^{(2)}(Z,N)$ for $A>80$ can be 
explained from the relation (\ref{eq:5}). ( Strictly speaking, there is a 
deviation from $40/A$ due to the mass dependence of $I_{0}$.) Furthermore, it 
is obtained that $\delta V^{(m)}(Z,N)$ for $m=3-6$ have similar pattern to 
$V^{(2)}(Z,N)$, and are also traced by the curve $40/A$. 

To analyze the double differences of binding energies, we now express the 
ground-state energy as follows:
\begin{eqnarray}
  E(Z,N)=<H>=E_{\rm sp}+E_{\pi\nu}^{P_{0}+QQ+P_{2}}+E_{\pi\nu}^{\tau=0}+
  E_{\pi\pi+\nu\nu}, \label{eq:18} \\
  E_{\rm sp}=\langle H_{\rm sp}\rangle, \hspace{3.5cm} \label{eq:19} \\
  E_{\pi\nu}^{P_{0}+QQ+P_{2}}=\langle V_{\pi\nu}^{P_{0}+QQ+P_{2}}\rangle, 
  \hspace{0.8cm} \label{eq:20} \\
  E_{\pi\nu}^{\tau=0}=\langle V_{\pi\nu}^{\tau=0}\rangle, \hspace{3cm} 
  \label{eq:21} \\
  E_{\pi\pi+\nu\nu}^{P_{0}+QQ+P_{2}}=\langle 
  V_{\pi\pi+\nu\nu}^{P_{0}+QQ+P_{2}}\rangle,  \hspace{0.8cm} \label{eq:22}
\end{eqnarray}
where $\langle \hspace{0.5cm} \rangle$ denotes the expectation value with 
respect to the ground state. Here, $V_{\pi\nu}^{P_{0}+QQ+P_{2}}$ is the 
$p-n$ parts of the $P_{0}+QQ+P_{2}$ force, and 
$V_{\pi\pi+\nu\nu}^{P_{0}+QQ+P_{2}}$ is the proton-proton ($p-p$) and 
neutron-neutron ($n-n$) parts of the total interaction (\ref{eq:8}). 
Figure 2 shows $E_{\pi\nu}^{P_{0}+QQ+P_{2}}$ and $E_{\pi\nu}^{\tau=0}$ as a 
function of valence-neutron number $n_{n}$ for Nb and Mo isotopes. The $p-n$ 
part of  the $P_{0}+QQ+P_{2}$ energy, $E_{\pi\nu}^{P_{0}+QQ+P_{2}}$, exhibits 
a characteristic odd-even staggering in Nb isotopes, while 
$E_{\pi\nu}^{\tau=0}$ gives a smooth line except for $n_{n}=1$. On the other 
hand, for Mo isotopes $E_{\pi\nu}^{P_{0}+QQ+P_{2}}$ varies smoothly as 
$n_{n}$ increases, and indicates very different structure from that of Nb 
isotopes. This can be attributed to extra energy for the odd-odd nuclei, which 
mainly comes from the $\tau=1$ $p-n$ part of the $P_{0}+QQ+P_{2}$ force. 

Consider the double difference of ground-state energies, $\delta^{(1)}E(Z,N)$, 
using the operator $\delta^{(1)}$ defined by Eq. (\ref{eq:102}). 
Since there are almost no contributions from the single particle energy 
$E_{\rm sp}$ and Coulomb interaction to the double difference of binding 
energies as seen in the form of Eq. (\ref{eq:102}), one  notices that 
$\delta^{(1)}E(Z,N)$ is able to be compared directly with the experimental 
value $\delta V^{(1)}(Z,N)$. 
 Figure 3 shows the calculated and experimental double differences of binding 
 energies, $\delta^{(1)}E(Z,N)$ and $\delta V^{(1)}(Z,N)$, as a function of 
 mass $A=N+Z$ for the Nb, Mo, Tc, and Pd isotopes. 
The Nb and Mo nuclei near $N=Z$ at the beginning of $g_{9/2}$ shell region 
probably have the mixing of single particle levels, the $p_{1/2}, f_{5/2}$, and
 $p_{3/2}$. As seen from Fig. 3(a)-(d), however, 
 the agreement with experiments 
is quite good. Our calculation reproduces the staggering, and also predicts 
the highest spikes at $N=Z$ nuclei though no experimental data are present. 
\begin{small}
\begin{description}
\item{TABLE I.} {The components of $\delta^{(1)}E(Z,N)$ for the Mo isotopes. The 
first and second columns denote the $\tau=1$ components, and the third and 
fourth columns the $\tau=0$ components.
}
\end{description}
\begin{center}
\begin{tabular}{|c|cc|cc|c|}\hline
     &  $\tau=1$  &   &  $\tau=0$ &   & Total \\ \hline
 $N$ & $\delta^{(1)}E_{\pi\nu}^{P_{0}+QQ+P_{2}}$ & 
 $\delta^{(1)}E_{\pi\pi+\nu\nu}$ & $\delta^{(1)}E_{\pi\nu}^{QQ}$ 
 & $\delta^{(1)}E_{\pi\nu}^{\tau=0}$ & $\delta^{(1)}E$ \\ \hline
 42  &  2.049 &  0.001 & -0.001 & -0.001 &  2.050 \\
 43  & -0.602 & -0.288 &  0.451 &  0.464 &  0.024 \\
 44  &  0.414 &  0.398 & -0.232 &  0.460 &  1.041 \\
 45  & -0.511 & -0.096 &  0.236 &  0.461 &  0.089 \\
 46  &  0.346 &  0.218 & -0.274 &  0.457 &  0.747 \\
 47  & -0.217 & -0.004 & -0.125 &  0.463 &  0.117 \\
 48  &  0.018 &  0.177 & -0.022 &  0.465 &  0.638 \\
 49  & -0.233 &  0.000 & -0.186 &  0.461 &  0.047 \\
 50  & -0.014 &  0.183 & -0.022 &  0.465 &  0.611 \\  \hline
\end{tabular}
\end{center}
\end{small}

Let us now analyze the staggering and the highest spikes at $N=Z$ .  
In Table I, the components of $\delta^{(1)}E(Z,N)$ for the Mo isotopes are listed. 
The components $\delta^{(1)}E_{\pi\nu}^{P_{0}+QQ+P_{2}}$, 
$\delta^{(1)}E_{\pi\nu}^{\tau=0}$ and $\delta^{(1)}E_{\pi\pi+\nu\nu}$ are 
obtained using the definition (\ref{eq:102}) of the operator $\delta^{(1)}$ 
for the respective parts of the ground state energies, 
Eqs. (\ref{eq:20})-(\ref{eq:22}). It is seen that the large value at $N=Z=42$ 
comes from only the $\tau=1$ component of 
$\delta^{(1)}E_{\pi\nu}^{P_{0}+QQ+P_{2}}$, and others are very small. Thus it 
is clear that the $\tau=1$ $p-n$ interaction of the $P_{0}+QQ+P_{2}$ force is 
closely related to the large values of $\delta V^{(1)}$ at $N=Z$. On the other 
hand, $\delta^{(1)}E(Z,N)$ for $N \neq Z$ exhibits staggering as seen in Table 
I [also see Fig. 3(b)]. The small values for odd $N$ are due to the 
cancellation of $\tau=1$ and $\tau=0$ components of $\delta^{(1)}E(Z,N)$, and 
for even $N$  both $\tau=1$ and $\tau=0$ components contribute in phase. 
The value of $\delta^{(1)}E^{\tau =0}_{\pi \nu}$ is 0 for $N=Z$ and 
${\frac{1}{2}}k^0$ for $N>Z$, though the tabulated values have numerical 
errors.

Figure 4 shows $\delta^{(2)}E(Z,N)$ and $\delta V^{(2)}(Z,N)$ as a function of 
mass $A=N+Z$ for the Mo, Tc, Pd, and Sn isotopes. The values of 
$\delta^{(2)}E(Z,N)$ are a little bit smaller than the experimental ones but 
the agreement is quite well. 
Our calculation predicts large $\delta^{(2)}E(Z,N)$ at $N=Z$. The components 
of $\delta^{(2)}E(Z,N)$ are shown for the Mo isotopes in Table II. 
\begin{small}
\begin{description}
\item{TABLE II.} {The components of $\delta^{(2)}E(Z,N)$ for the Mo isotope. }
\end{description}
\begin{center}
\begin{tabular}{|c|cc|cc|c|}\hline
     &   $\tau=0$  &   &   $\tau=1$  &  &  Total \\ \hline
  $N$ & $\delta^{(2)}E_{\pi\nu}^{QQ}$ & 
  $\delta^{(2)}E_{\pi\nu}^{\tau=0}$ &  
  $\delta^{(2)}E_{\pi\nu}^{P_{0}+QQ+P_{2}}$ & $\delta^{(2)}E_{\pi\pi+\nu\nu}$ 
  & $\delta^{(2)}E$ \\ \hline
 42  &   0.275 &  0.694 &  0.365 &  -0.295 &  1.039 \\
 43  &   0.237 &  0.578 &  0.128 &  -0.125 &  0.819 \\
 44  &   0.110 &  0.462 & -0.094 &   0.058 &  0.536 \\
 45  &  -0.006 &  0.462 & -0.083 &   0.069 &  0.441 \\
 46  &  -0.021 &  0.461 & -0.083 &   0.063 &  0.419 \\
 47  &  -0.042 &  0.461 & -0.092 &   0.070 &  0.397 \\
 48  &  -0.072 &  0.463 & -0.099 &   0.080 &  0.373 \\
 49  &  -0.088 &  0.463 & -0.110 &   0.089 &  0.355 \\
 50  &  -0.105 &  0.464 & -0.123 &   0.094 &  0.331 \\  \hline
\end {tabular}
\end{center}
\end{small}
It is seen that 
$\delta^{(2)}E(Z,N)$ comes from only the $\tau=0$ $p-n$ interaction, and the 
$\tau=1$ components are small because of the cancellation of 
$\delta^{(2)}E_{\pi\nu}^{P_{0}+QQ+P_{2}}(Z,N)$ and 
$\delta^{(2)}E_{\pi\pi+\nu\nu}(Z,N)$. 
The $\tau=0$ component of $QQ$ is small except for $N=42$ and 
43. (In other isotopes, this component is small for all $N$, then this 
behavior for $N=42,43$ in the Mo isotopes is exceptional.)
The value of $\delta^{(2)}E^{\tau =0}_{\pi \nu}$ is $3k^0/4$ 
 for $N$=$Z$, $5k^0/8$ for $|N-Z|=1$ and $k^0/2$
 for $|N-Z|>1$, though the tabulated values have numerical errors.
Thus $\delta V^{(2)}(Z,N)$ derived from the experimental binding energies are 
considered to be nearly attributed to the $\tau=0$ $p-n$ force 
$V_{\pi\nu}^{\tau=0}$.
Namely, we deduce an approximate relation
\begin{eqnarray}
 \delta V^{(2)}(Z,N) & \approx & \delta^{(2)}E(Z,N) \nonumber \\
  & \approx & \delta^{(2)}E^{\tau =0}_{\pi \nu}(Z,N).  \label{eq:23}
\end{eqnarray}
 This is consistent with the argument that $\delta V^{(2)}(Z,N)$ vanishes if 
 one neglects the $\tau=0$  $p-n$ interaction in the shell model calculation 
 with a surface $\delta$ interaction for the $2s-1d$ shell \cite{Brenner}. 

Furthermore, we calculated $\delta^{(1)}E(Z,N)$ and $\delta^{(2)}E(Z,N)$ for 
the $1f_{7/2}$ shell nuclei using the shell model
 ($f_{7/2}$)$^n$ with the force parameters (\ref{eq:110}). 
Figure 5 shows  the calculated and experimental double differences of binding 
energies for the Ti and Cr isotopes. The agreement with experiments is very good. 
Our effective interaction (${P_{0}+QQ+P_{2}+V_{\pi\nu}^{\tau=0}}$) 
reproduces well the experimental values of the double differences of binding 
energies also in the $1f_{7/2}$ shell region .
 This supports our model Hamiltonian being
 applicable to a wide range of nuclei. The good agreement tells us that the 
 results in the $g_{9/2}$ shell region are reliable and  give good predictions.
It should be also noted that the approximate relation Eq. (\ref{eq:23}) holds 
in this 
 $1f_{7/2}$ shell region.

  It is now meaningful that the $p-n$ correlation energy $E^{\tau=0}_{\pi \nu}$
  is expressed as 
\begin{eqnarray}
 E^{\tau=0}_{\pi \nu}=- \frac{1}{2}k^{0} \{ \frac{n}{2}(\frac{n}{2}+1)-
 T(T+1)\} ,  \label{eq:2000}
\end{eqnarray}
for states with the total valence-nucleon number ${n}$ and
 total isospin ${T}$ from Eq. (\ref{eq:11}).
 For $N-Z>1$, we can easily show that
\begin{equation}
 \delta^{(2)}E_{\pi\nu}^{\tau=0} = \frac{k^{0}}{2}. \label{eq:25}
\end{equation}
The global behavior of $\delta V^{(2)}(Z,N)$ depending on $40/A$ 
as seen in Fig. 1(b) combined with the relations (\ref{eq:23}) and 
(\ref{eq:25})
suggests that in a wide-range view the force strength $k^0$ might have $1/A$ 
dependence
\begin{equation}
 k^0 \approx \frac{80}{A}. \nonumber
\end{equation}
(Strictly speaking, since the double difference 
$\delta^{(2)}E_{\pi\nu}^{\tau=0}$ with $k^0 = 80/A$ deviates from the curve 
$40/A$, we need a higher-order correction with $1/A^{4/3}$ to reproduce the 
curve $40/A$.) 
In fact, the force parameters $k^0$ employed, 0.925 MeV for the 
$1g_{9/2}$ shell
 nuclei and 1.9 MeV for the $1f_{7/2}$ shell nuclei, reflect some 
 $A$ dependence.
 These values do not very deviate from the global value $80/A$, if we compare 
 them
 with the examples $80/A=0.93$ for $A=86$ and $80/A=1.74$ for $A=46$. 
 Certainly, if we impose the $1/A$ dependence on $k^0$ like $1.9 \times (48/A)$
 in the calculations for the $1f_{7/2}$-shell nuclei, the binding 
 energies obtained
 for $N \approx Z$ nuclei are reproduced better \cite{Hasegawa}.  This 
 improves the double difference
 of binding energies $\delta E^{(2)}(Z,N)$ as seen in Fig. 5,
  where the circles stand for the constant $k^0$ and the crosses for the
  $k^0=1.9 \times (48/A)$.
  Accordingly, the observed variation $40/A$ in $\delta V^{(2)}(Z,N)$ is 
  suggested
  to be mainly attributed to the global dependence $80/A$ on $k^0$.
The $\tau=0$ $p-n$ force $V^{\tau =0}_{\pi \nu}$ is possibly applicable to
 $N>Z$ nuclei with $T=T_z=(N-Z)/2$.  If it is true, the relation (\ref{eq:25})
 holds for $N>Z$ too.  
 The $p-n$ force $V^{\tau =0}_{\pi \nu}$ with the global
 parameter $k^0=80/A$ (with correction) may explain the smooth systematic 
 behavior of
  $\delta V^{(2)}(Z,N)$ in the mass region $A>80$ as seen in Fig. 1(b).
 This must be examined further.
 So far, we have not adopted the $A$ dependence for the $P_0$, $QQ$, 
 and $P_2$ forces,
 because we do not have any strong demand to do so within the present 
 calculations in a very tiny model space using a single $j$ shell.
 It must, however, be necessary for our model when
 we make quantitative calculations in many $j$ shells.

\section{Symmetry energy and Wigner energy}

Let us next discuss the symmetry energy $E_{\rm sym}=a_{\rm sym}(N-Z)^{2}/A$ 
and 
Wigner energy $E_{W}$ using the same set of parameters as the in 
previous sections. 
The experimental data indicate that the symmetry energy accompanied by the 
Wigner energy is proportional to the $T(T+1)$ where $T=T_{z}=|N-Z|/2$. Since 
$E_{\pi\nu}^{\tau=0}$ includes the $T(T+1)$ term as 
seen in Eq. (\ref{eq:2000}), both quantities must be closely related to the isoscalar $p-n$ 
force $V_{\pi\nu}^{\tau=0}$.  Figure 6 shows the symmetry energy 
coefficient $a(A)=4a_{\rm sym}$ in the expression 
$E_{\rm sym}+E_{W}=a(A)T(T+1)/A$ for the $f_{7/2}$ shell nuclei. The symmetry 
energy coefficient can be extracted by the treatment of 
J$\ddot{\rm a}$necke and Comay \cite{Janecke,Comay}.
We calculated the Coulomb-energy-corrected binding 
energies $B^{*}=B({\rm exp})+E_{\rm Coul}({\rm cal})$ following 
Caurier $et al.$ \cite{Caurier}.
  The calculated symmetry 
energy coefficients nicely reproduce the experimental data in Fig. 6. Where 
does the symmetry energy come from? We should now analyze the result obtained. 
If $V(QQ)$ and $V(P_{2})$ are eliminated from the total Hamiltonian, 
the Hamiltonian $H_{\rm sp}+V(P_{0})+V_{\pi \nu}^{\tau=0}$ has SO(5) symmetry 
in the single-$j$ shell approximation. 
The total energy of $H_{\rm sp}+V(P_{0})+V_{\pi\nu}^{\tau=0}$ in the 
single-$j$ shell is 
specified by the total valence nucleon number $n=n_{p}+n_{n}$ and the 
total isospin $T$ as follows \cite{Hecht}:
\begin{eqnarray}
 \tilde{E}= E_{\rm sp} + E_{P_{0}} + E_{\pi\nu}^{\tau=0}  \hspace{9cm} \nonumber \\
  = \epsilon n - \frac{1}{2}\{ g_{0}n(\Omega - \frac{n-6}{4}) + 
  k^{0}\frac{n}{2}(\frac{n}{2}+1) \} + \frac{1}{2}(g_{0}+k^{0})T(T+1), 
  \label{eq:26} 
\end{eqnarray}
where $E_{\rm sp}$, $E_{P_{0}}$ and $E_{\pi\nu}^{\tau=0}$ denote the expectation values of 
$H_{\rm sp}$, $V(P_{0})$ and $V_{\pi\nu}^{\tau=0}$ with respect to ground states 
with $n$ and $T$, respectively. 
From the coefficient of the $T(T+1)$ part in Eq. 
(\ref{eq:26}), the symmetry 
energy coefficient $\tilde{a}(A)$ is expressed as 
\begin{equation}
 \tilde{a}(A)= \frac{1}{2}(g_{0}+k^{0})A, \label{eq:27}
\end{equation}
which is proportional to the sum of the force strengths $g_{0}$ and $k^{0}$. 
The parameter set (\ref{eq:100}) gives the value
 $\tilde{a}(A)=1.245A$ in the $f_{7/2}$ region. As shown in Fig. 6, the 
 symmetry energy coefficient $\tilde{a}(A)$ almost describes that obtained with the 
 total Hamiltonian including $V(QQ)$ and $V(P_{2})$. Thus, we see that the 
 symmetry energy in this region originates in the isoscalar 
 $p-n$ force $V_{\pi\nu}^{\tau=0}$ and 
 isovector $J=0$ pairing force. Their contributions are 76$\%$ and 24$\%$, 
 respectively, in the present calculation. 

In the expression $E_{\rm sym}+E_{W}=a(A)T(T+1)/A$, the Wigner energy has the same 
coefficient as the symmetry energy, and is expressed as $E_{W}=a(A)T/A$. 
Poves and Mart$\acute{\rm i}$nez-Pinedo pointed out that a shell model calculation with the 
$J=0$ isovector pairing force and $J=1$ isoscalar pairing force in the $N \approx Z$  $fp$ shell nuclei cannot explain the magnitude of the experimental 
Wigner energy \cite{Poves}. If we take the same parameter $g_{0}=0.295$ for the 
$J=0$ isovector pairing force as that of Ref.\cite{Poves}, the Wigner energy 
is estimated as $E_{W}=g_{0}|N-Z|/4=3.54|N-Z|/A$ MeV for $A=48$. This value is 
not very different from the result $E_{W}=3.04|N-Z|/A$ MeV they obtained. 
The empirical Wigner energy, $E_{W}=47|N-Z|/A$ MeV \cite{Satula2} or 
$37|N-Z|/A$ MeV \cite{Vogel1,Duflo} is very large compared with these values. 
Figure 6 tells that the Wigner energy cannot be reproduced without 
$V_{\pi\nu}^{\tau=0}$. 
 If we introduce $V_{\pi\nu}^{\tau=0}$, the Wigner energy becomes 
 $E_{W}=37.4|N-Z|/A$ MeV for $A=60$ from Eq. (\ref{eq:27}), 
 which is consistent 
 with the empirical formula. We can conclude that the isoscalar $p-n$ force 
 and isovector $J=0$ pairing force are origin of both the symmetry energy and 
 Wigner energy. In particular, it is important to note 
 that the isoscalar $p-n$ 
 interaction with all $J$ components, not only $J=1$, 
 is crucial for reproducing 
 the symmetry energy and the Wigner energy. 

\section{Odd-even mass difference}

The odd-even mass difference in three-point and four-point 
expressions,
 \begin{eqnarray}
  & & \Delta_{3}(Z,N)=\frac{(-1)^{N}}{2}[B(Z,N+1)-2B(Z,N)+B(Z,N-1)], 
  \label{eq:201} \\
  & & \Delta_{4}(Z,N)=\frac{(-1)^{N}}{4}[B(Z,N+1)-3B(Z,N)+3B(Z,N-1) 
  \nonumber \\
  & & \hspace{8cm} -B(Z,N-2)], \label{eq:202}
\end{eqnarray}
 is often used to estimate the empirical pairing gap (for neutron) and to 
 determine the pairing force strength. 
Figure 7 shows the experimental values of $\Delta_{3}(Z,N)$ and 
$\Delta_{4}(Z,N)$ as a function of $N-Z$ in even-$Z$ isotopes with proton 
number $Z$=20-30 and 36. In Fig. 7(a), $\Delta_{3}(Z,N)$ exhibits staggering 
around 1.5MeV and has a notable peak at $N=Z$, while 
$\Delta_{4}(Z,N)$ has a hill near 
$N=Z$ and $N=Z+1$ but varies smoothly for $ N\geq Z+2$. 
The OEMD, $\Delta_{3}(Z,N)$ and $\Delta_{4}(Z,N)$, are about 1.5 MeV on the average for $N\geq Z+2$. This value is usually regarded as a measure of the empirical neutron pairing gap. 
In addition, 
we notice the asymmetry of $\Delta_{3}(Z,N)$ with respect to $N-Z=0$. 
This may be due to the so-called Nolen-Schiffer anomaly \cite{Nolen-Schiffer}, 
an energy difference between neighboring mirror nuclei, 
which cannot be explained 
by the electromagnetic interaction.

The calculated values of $\Delta_{3}(Z,N)$ and
 $\Delta_{4}(Z,N)$ are obtained by replacing $B(Z,N)$ by 
 the ground-state energy $E(Z,N)$ in Eqs. (\ref{eq:201}) and 
 (\ref{eq:202}), since the
 Coulomb energy hardly contributes to $\Delta_{3}(Z,N)$ and
 $\Delta_{4}(Z,N)$. 
 In our single $j$ shell model, the single-particle energy has no contribution 
 to $\Delta_{3}(Z,N)$ and $\Delta_{4}(Z,N)$.  
Figure 8 shows the calculated and experimental values of $\Delta_{3} (Z,N)$ as 
a function of $N-Z$ for the Ca, Ti, Cr, and Fe isotopes. 
The agreement with experiments is quite good. Especially, the observed peaks 
at  $N=Z$ are well reproduced. 
\begin{small}
\begin{description}
\item{TABLE III.} {The components of $\Delta_{3} (Z,N)$ for Cr the isotopes. The 
first and second columns denote the $\tau=1$ components, and the third and 
fourth columns the $\tau=0$ components.}
\end{description}
\begin{center}
\begin{tabular}{|c|cc|cc|c|}\hline
     &  $\tau=1$  &  &  $\tau=0$ &  & Total \\ \hline
 $N$ & $\Delta_{\pi\nu}^{P_{0}+QQ+P_{2}}$ & 
 $\Delta_{\pi\pi+\nu\nu}^{P_{0}+QQ+P_{2}}$ & 
 $\Delta_{\pi\nu}^{QQ}$ & $\Delta_{\pi\nu}^{\tau=0}$ & 
 $\Delta_{3} (Z,N)$ \\ \hline
 21  &  0.001 &  1.553 &  0.002 &  0.000 &  1.556 \\
 22  & -0.027 &  1.984 &  0.106 & -0.004 &  2.059 \\
 23  &  0.008 &  1.657 & -0.107 & -0.003 &  1.555 \\
 24  &  0.750 &  1.500 & -0.009 &  0.469 &  2.710 \\
 25  &  0.003 &  1.662 & -0.111 & -0.013 &  1.541 \\
 26  & -0.034 &  2.007 &  0.089 & -0.015 &  2.047 \\
 27  &  0.001 &  1.579 & -0.015 & -0.005 &  1.560 \\ \hline
\end {tabular}
\end{center}
\end{small}

 Let us analyze what interactions are important in $\Delta_{3}(Z,N)$.
 We separately calculated the contributions of the interaction energies
 $E_{\pi \nu}^{P_0+QQ+P_2}$, $E_{\pi \nu}^{\tau =0}$ and
 $E_{\pi \pi + \nu \nu}^{P_0+QQ+P_2}$ to $\Delta_{3}(Z,N)$,
 and denote them by
 $\Delta_{\pi \nu}^{P_0+QQ+P_2}$, $\Delta_{\pi\nu}^{\tau=0}$ and
 $\Delta_{\pi\pi+\nu\nu}^{P_0+QQ+P_2}$, shortly. 
In Table III, the components of $\Delta_{3} (Z,N)$ are listed for the Cr isotopes. 
This table indicates the dominance of
 $\Delta_{\pi\pi+\nu\nu}^{P_0+QQ+P_2}$, namely the dominance
 of the like-nucleon pairing correlations. 
The other components of $\Delta_{3}(Z,N)$ are very small except that the 
isovector parts of $\Delta_{\pi\nu}^{P_{0}+QQ+P_{2}}$ and 
$\Delta_{\pi\nu}^{\tau=0}$ have large values at $N=Z$. 
We see their additional contributions for the large peaks of
 $\Delta_{3}(Z,N)$ at $N=Z$ in Fig. 8.
 The isovector and isoscalar $p-n$ interactions are most
 cooperative with the $p-p$ and $n-n$ interactions in the $N=Z$ nuclei.

 This situation is explained by illustrating the behavior of the respective 
 interaction energies in Fig. 9.  The staggering of $\Delta_{3}(Z,N)$ in Fig. 7
 is almost attributed to that of $E_{\pi \pi + \nu \nu}^{P_0+QQ+P_2}$ in
 Fig. 9.  The straight lines of the interaction energies
 $E_{\pi \nu}^{P_0+QQ+P_2}$ and $E_{\pi \nu}^{\tau =0}$ go down
 as $N$ increases and turn to the different directions at $N=Z$.
 The coincident "bends" at $N=Z$ cause the increase
 (the peak) of $\Delta_{3}(Z,N)$ according to the form of
 Eq. (\ref{eq:201}).
 These bends of the $p-n$ interaction energies produce the increase of
 $\delta V^{(2)}$ around $N=Z$.
 The bends give a special energy gain to the $N=Z$ even-even
 nuclei, $^{44}$Ti, $^{48}$Cr, $^{52}$Fe etc.  The $\alpha$-like
 four-nucleon correlations in these $N=Z$ nuclei can be interpreted in terms of
 the characteristic behavior of the $p-n$ interactions in cooperation with
 the like-nucleon interactions \cite{Hasegawa1}.

\begin{small}
\begin{description}
\item{TABLE IV.} {The components of $\Delta_{\pi\pi+\nu\nu}^{P_{0}+QQ+P_{2}}$ 
for the Cr isotopes.}
\end{description}
\begin{center}
\begin{tabular}{|c|ccc|c|}\hline
 $N$ & $\Delta_{\pi\pi+\nu\nu}^{P_{0}}$ & $\Delta_{\pi\pi+\nu\nu}^{QQ}$ & 
 $\Delta_{\pi\pi+\nu\nu}^{P_{2}}$ & $\Delta_{\pi\pi+\nu\nu}^{P_{0}+QQ+P_{2}}$ 
 \\ \hline
 41  &  1.182 &  0.374 & -0.003 &  1.553 \\
 42  &  1.919 &  0.589 & -0.524 &  1.984 \\
 43  &  2.120 &  0.279 & -0.742 &  1.657 \\
 44  &  1.901 &  0.398 & -0.799 &  1.500 \\
 45  &  2.129 &  0.281 & -0.748 &  1.662 \\
 46  &  1.950 &  0.594 & -0.537 &  2.007 \\
 47  &  1.211 &  0.381 & -0.013 &  1.579 \\ \hline
\end {tabular}
\end{center}
\end{small}
  According to Refs. \cite{Hakkinen,Satula3}, on the other hand,
 the OEMD in light nuclei is strongly affected by deformation
 originated in the Jahn-Teller mechanism \cite{Nazarewicz}.
 It is interesting to see what interactions contribute to 
 $\Delta_{\pi\pi+\nu\nu}^{P_0+QQ+P_2}$ being the main part of
 $\Delta_3$.
 Table IV presents the respective contributions of the $P_0$,
 $QQ$ and $P_2$ forces to
 $\Delta_{\pi\pi+\nu\nu}^{P_0+QQ+P_2}$ in the Cr isotopes. 
 The dominant component is $\Delta_{\pi\pi+\nu\nu}^{P_0}$
 as expected, i.e., about 2.0 MeV for $\mid N-Z \mid \geq 2$ and
 about 1.2 MeV for $N=Z\pm 3$.
 In addition, there are considerably large contributions of
 $\Delta_{\pi\pi+\nu\nu}^{QQ}$ and $\Delta_{\pi\pi+\nu\nu}^{P_2}$.
 Since the $QQ$ correlation is intimately related to the nuclear
 deformation, the positive contribution of
 $\Delta_{\pi\pi+\nu\nu}^{QQ}$  is consistent with the conclusion
 of Ref. \cite{Satula3}. The contribution of quadrupole pairing
 force $P_2$, however, is negative, and is larger than that of the
 $QQ$ correlation for $\mid N-Z \mid \geq 2$.
 This is easily understood by the fact that the quadrupole pairing
 correlation breaks the $J=0$ Cooper pairs of neutrons.
 The present calculation tells us that the quadrupole pairing correlation
 probably cancels the effect of the $QQ$ correlation or the
 deformation on the OEMD value \cite{Hakkinen,Satula3}.

 Figure 10 shows calculated and experimental values of
 $\Delta_{4}(Z,N)$ as a function of $N-Z$ for the Ca, Ti, Cr, and Fe
 isotopes. The agreement with experiments is quite good except for
 the Ti isotopes. The calculation reproduces the hill
 near $N=Z$ and $Z+1$ and also the gentle behavior of
 $\Delta_{4}(Z,N)$ near the value 1.6 MeV in the region of $N>Z+1$
 and $N<Z$. 
 The increase of $\Delta_{4}(Z,N)$ at $N=Z$ and $N=Z+1$ is explained
 in terms of the same mechanism as that of $\Delta_3(Z,N)$ at $N=Z$,
 which is caused by the coincident bends at $N=Z$ of the two graphs in
 Fig. 9 that illustrate the variations of the $p-n$ interaction energies.
 The isovector p-n interactions of the $P_0+QQ+P_2$ force
 and the isoscalar $p-n$ force $V_{\pi \nu}^{\tau =0}$ are important
 at $N=Z$ also for $\Delta_{4}(Z,N)$.

\section{Two-proton separation energy}

     We have investigated several quantities related to the nuclear
 binding energy in the previous sections. The calculated data include
 binding energies of nuclei close to the proton drip line. It is
 interesting to look at the two-proton separation energy, experimental
 data of which has been accumulated by the radioactive beam.
 It provides the possibility for studying new decay modes such as
 diproton emission. Some nuclei around $^{48}$Ni are expected to
 possibly be two-proton emitters. There is a large deviation between
 theory and experiment for the two-proton separation energy up to now.
 All the predictions by the Hartree-Fock (Bogoliubov) and relativistic
 Hartree-Fock (Bogoliubov) treatments \cite{Nazare,Lalazissis} underestimate 
 the two-proton separation energies at $N=Z$ nuclei. This discrepancy could
 be due to the lack of p-n interactions in these treatments.
 As seen in the OEMD in Sec. V, the p-n correlations cooperate
 with the $p-p$ and $n-n$ correlations especially at $N=Z$ nuclei.
 We can expect that the $p-n$ interactions have a considerable influence
 on the two-proton separation energy.
 
  We calculated the two-proton  separation energies $S_{2p}$
 for the  $f_{7/2}$ shell nuclei with $Z=20-28$ and $N=20-28$.
  In the calculation, the force strength $k^0$
 is chosen as $k^0=1.9 \times (48/A)$ which was used in the previous
 paper \cite{Hasegawa}, because the 1/A dependence of $k^0$ improves the binding
 energy and is also supported by the discussion in Sec. III.
 Calculated values of $S_{2p}$ are compared with experimental data
 taken from Ref. \cite{Table} in Fig. 11.
 Here we subtracted the Coulomb energy following Caurier $et al.$
  \cite{Caurier}.
  The agreement is good. The observed
 values of $S_{2\rho}$ are reproduced well at $N=Z$ nuclei.
 Again, the isoscalar $p-n$ force $V^{\tau=0}_{\pi\nu}$ plays
 an important role in the two-proton separation energy at $N=Z$
 as well as in the other quantities discussed above.
 The model space $(f_{7/2})^n$ and the set of parameters used are
 not appropriate for nuclei with large $Z$ and $N$, strictly speaking.
 According to the experience in this paper, however, the results on
 the {\it different quantities of binding energies} might be still
 meaningful.  We note the prediction of our calculation that
 $^{48}$Ni, $^{47}$Co and $^{49}$Ni are possibly unstable,
 $^{50}$Ni is bound, and $S_{2p}$ of $^{46}$Fe and $^{48}$Co
 are close to zero.

\section{Conclusion}

 We have studied the $p-n$ interactions using the functional effective
 interaction with four force parameters which reproduces the
 energy levels and binding energies of $N \approx Z$ nuclei
 considerably well.
 
 First, we analyzed the double differences of binding energies
 $\delta V^{(1)}$ and $\delta V^{(2)}$, because the two quantities are
 expected to directly represent the $p-n$ interactions.
 Our effective interaction reproduces fairly well the experimental
 values of $\delta V^{(1)}$ and $\delta V^{(2)}$, and their
 characteristic behaviors, in the $g_{9/2}$ and $f_{7/2}$ shell
 nuclei.  The staggering of $\delta V^{(1)}$ is due to the
 competition between $\tau =1$ and $\tau =0$ components.  The large 
 spike of $\delta V^{(1)}$ at $N=Z$ is attributed to the $\tau =1$ $p-n$
 interactions of the $P_0+QQ+P_2$ foce, and 
 that of $\delta V^{(2)}$ at $N=Z$ contrarily represents the $\tau =0$ $p-n$
 interactions.  The observed values of $\delta V^{(2)}$ with respect to 
 mass $A$ are approximated by the curve $40/A$. This curve may be
 explained by granting an $A$ dependence on the $\tau =0$ p-n force
 strength $k^0$.
 
 Second, our effective interaction has also reproduced well the symmetry
 energy and the Wigner energy for the $f_{7/2}$ shell nuclei.
 The strong $\tau =0$ $p-n$ force $V_{\pi \nu}^{\tau =0}$ with assistance
 of the $\tau =1$, $J=0$ pairing force is important to explain 
 the magnitudes of the two quantities in this region.  
 It should be noted that the isospin parts
 of the two forces are proportional to $T(T+1)$ and their sum
 is directly related to the symmetry energy in the mass.
 The $A$ dependence of the symmetry energy
 coefficient seems to be determined mainly by that of $k^0$.

 Third, our effective interaction has described well the observed
 values of the odd-even mass difference ($\Delta_3$ and $\Delta_4$)
 for the $f_{7/2}$ shell nuclei.  The cooperation of the $p-n$
 interactions with the like-nucleon interactions is remarkable
 at $N=Z$.  It causes the rise of $\Delta_3$ and $\Delta_4$
 at $N=Z$.  The characteristic behaviors of the $p-n$ interaction
 energies at $N=Z$ (see Fig. 9) have an important effect not only on the 
 double differences
 of binding energies but in the odd-even mass difference.
 
 We have briefly touched on the two-proton separation energy $S_{2p}$
 using the calculated binding energies.  The calculation indicates
 a considerably large effect of the $p-n$ force
 on $S_{2p}$ at $N=Z$.
 We noted the prediction of our calculation for $S_{2p}$ near the
 $f_{7/2}$ proton drip line.
 
 The present investigations have shown that the $p-n$ interactions cause
 the notable behaviors of the observed quantities related to the binding
 energy near $N=Z$ in nuclei where valence protons and neutrons occupy
 the same shells.
 Furthermore, our calculations suggest that the $p-n$ interactions
  (especially $V_{\pi\nu}^{\tau =0}$) are important for describing these
  quantities over a wide range of $N>Z$ nuclei including the neutron drip line.
 
 All the results support the usefulness of the functional effective
 interaction composed of the four forces $P_0$, $QQ$, $P_2$ and
 $V_{\pi \nu}^{\tau =0}$, and clarify the essential roles of
 $V_{\pi \nu}^{\tau =0}$ in the physical quantities related to
 the binding energy.
 The present calculations, however, have been carried out
 in the single $j$ shell model.  Calculations in more realistic
 model spaces are in progress.
 
\begin{center}
{\bf Acknowledgments}
\end{center}

The authors are grateful to J.-Y. Zhang for helpful discussions. 

\newpage

\newpage
{\bf Figure captions}
\begin{description}
\item{Fig. 1} 
 Plots of the double differences of binding energies derived from the
 experimental masses in the region $A=16-164$:
 (a) $\delta V^{(1)}(Z,N)$ as a function of $A=N+Z$;
 (b) $\delta V^{(2)}(Z,N)$ as a function of $A=N+Z$.
 The dots stand for even-$A$ nuclei, and the crosses for odd-$A$
 nuclei. The curve $40/A$ is drawn both in (a) and (b).
\item{Fig. 2} 
 The calculated $p-n$ interaction energies as a function of
 the valence-neutron number $n_{\rm n}$ for (a) the Nb isotopes and
 (b) the Mo isotopes.  The open circles stand for the $p-n$ part of
 the interaction energy $E_{\pi\nu}^{P_0+QQ+P_2}$, the open squares
 for the $\tau=0$ $p-n$ interaction energy $E_{\pi\nu}^{\tau=0}$,
 and the diamonds for the total $p-n$ energy.
\item{Fig. 3} 
 The calculated values $\delta^{(1)}E(Z,N)$
 (open circles) and the experimental values
 $\delta^{(1)}V(Z,N)$ (solid squares)
 as a function of $A=N+Z$ for the Nb, Mo, Tc and Pd isotopes.
\item{Fig. 4} 
 The calculated values $\delta^{(2)}E(Z,N)$
 (open circles) and the experimental values
 $\delta^{(2)}V(Z,N)$ (solid squares)
 as a function of $A=N+Z$ for the Mo, Tc, Pd and Sn isotopes.
\item{Fig. 5} 
 The double differences of binding energies for the Ti and Cr isotopes,
 shown in the same manner as Figs. 3 and 4.
 The solid squares denote the experimental values,
 and the open circles and crosses denote the calculated values with
 $k^0=1.9$ and $k^0=1.9\times (48/A)$, respectively.
\item{Fig. 6} 
 The symmetry energy coefficients $a(A)$ in the $f_{7/2}$ shell region.
 The calculated and experimental values are denoted by the open circles and
 solid squares, respectively. The diamonds represent $\tilde{a}(A)=1.245A$
 in the $J=0$ isovector plus $J=$odd isoscalar pairing force model, and 
 the crosses $\tilde{a}(A)=0.295A$ in the $J=0$ isovector pairing force model. 
\item{Fig. 7} 
 Dependence of the experimental odd-even mass differences,
 $\Delta_3(Z,N)$ and $\Delta_4(Z,N)$ on $N-Z$ for nuclei
 with proton number $Z$=20-30 and 36.
\item{Fig. 8} 
 The calculated and experimental odd-even mass differences
 $\Delta_{3}(Z,N)$ as a function of $N-Z$ for the Ca, Ti, Cr and Fe
 isotopes.
\item{Fig. 9} 
 Interaction energies of the $P_0+QQ+P_2$ force and
  $V_{\pi \nu}^{\tau =0}$ in the Cr isotopes. 
\item{Fig. 10} 
 The calculated and experimental odd-even mass differences
 $\Delta_{4}(Z,N)$ as a function of $N-Z$ for the Ca, Ti, Cr and Fe
 isotopes.
\item{Fig. 11} 
 The two-proton separation energy $S_{2p}$ in the $f_{7/2}$ shell
 region. The calculated and experimental values are denoted by the
 open circles and solid squares, respectively. The force strength $k^{0}$
 is taken as $k^{0}=1.9\times (48/A)$.

\end{description}

\newpage
\begin{center}
     \epsfxsize=17cm
     \epsfysize=22cm
     \epsfbox{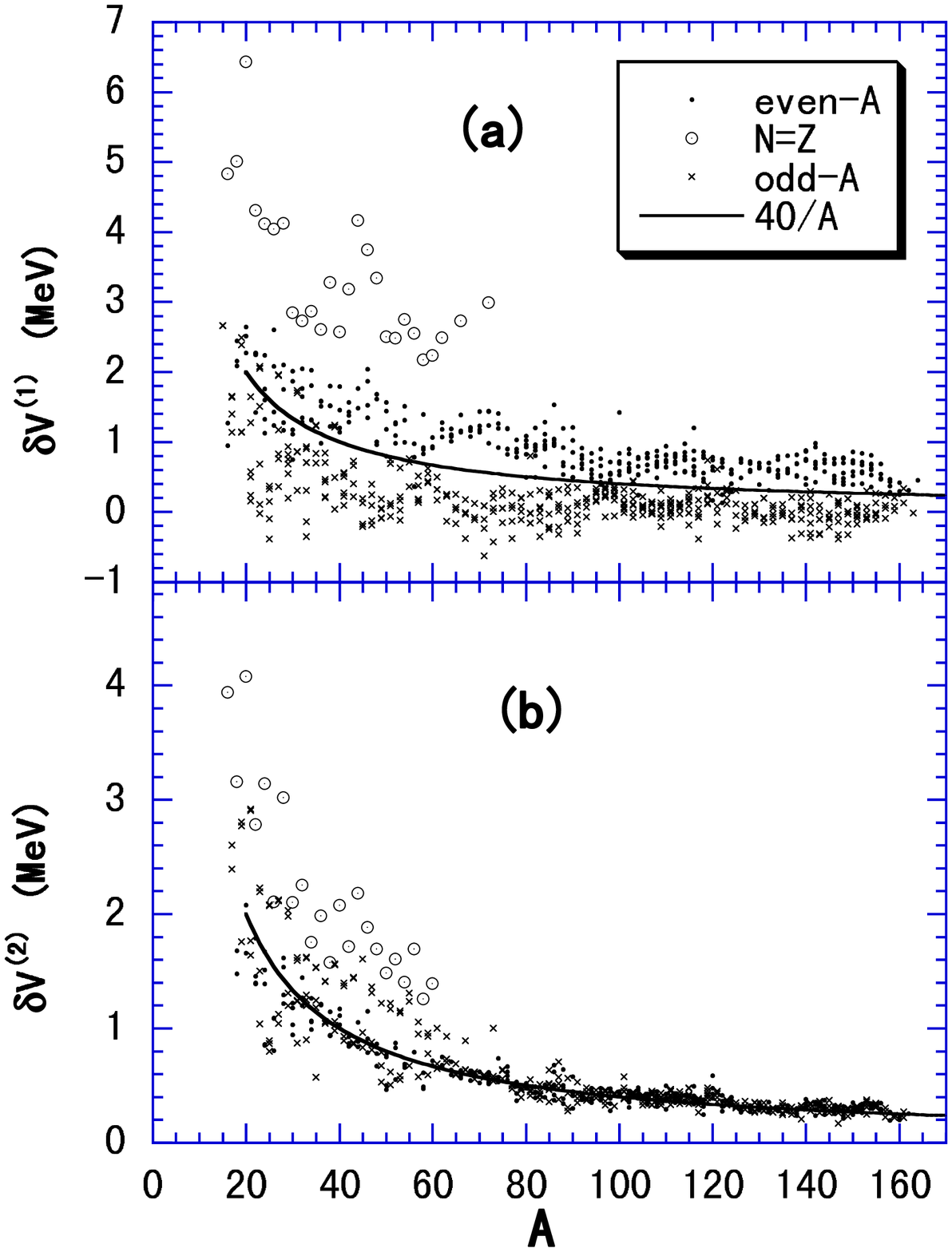}
\end{center}
\begin{center}
     \epsfxsize=17cm
     \epsfysize=22cm
     \epsfbox{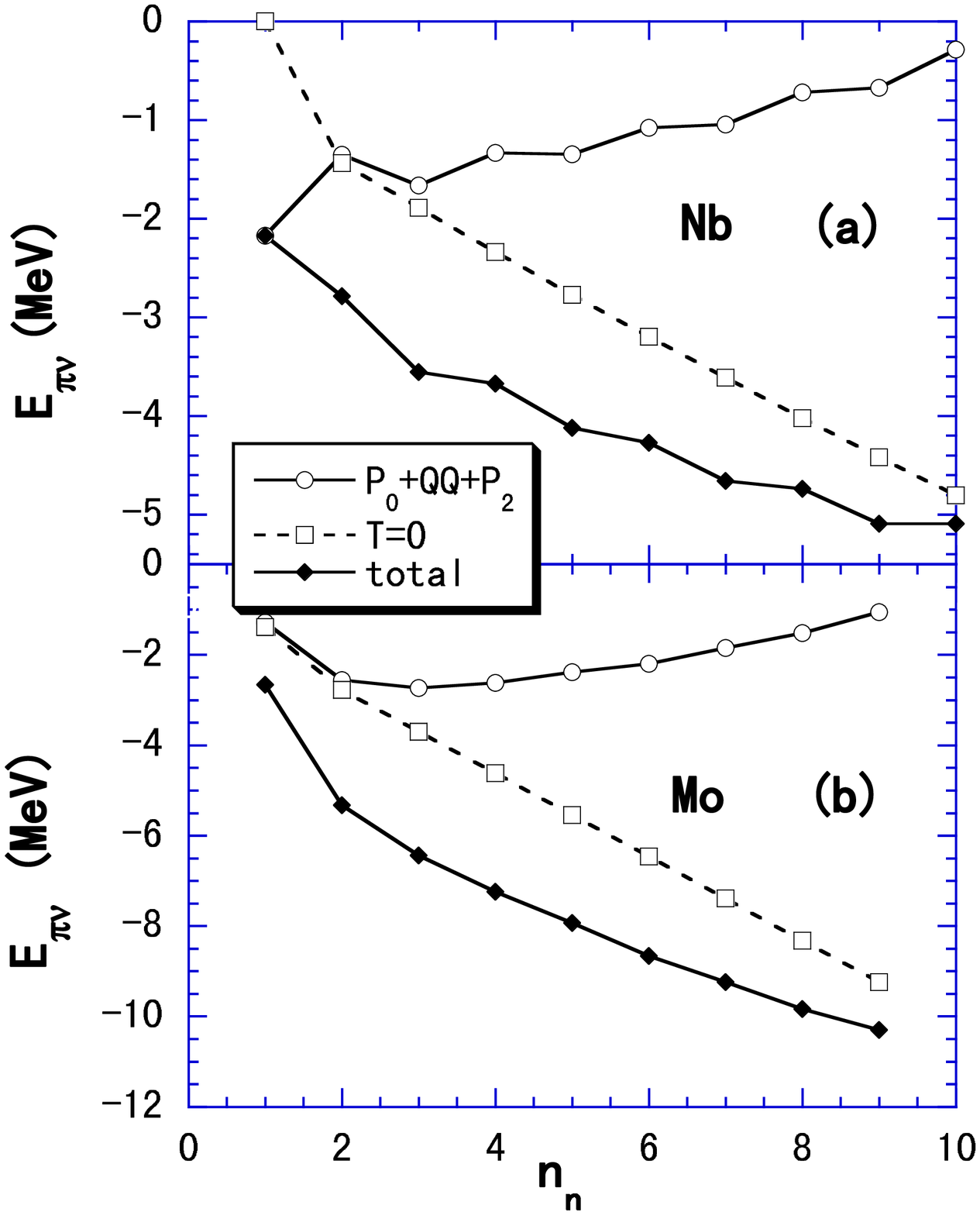}
\end{center}
\begin{center}
     \epsfxsize=17cm
     \epsfysize=22cm
     \epsfbox{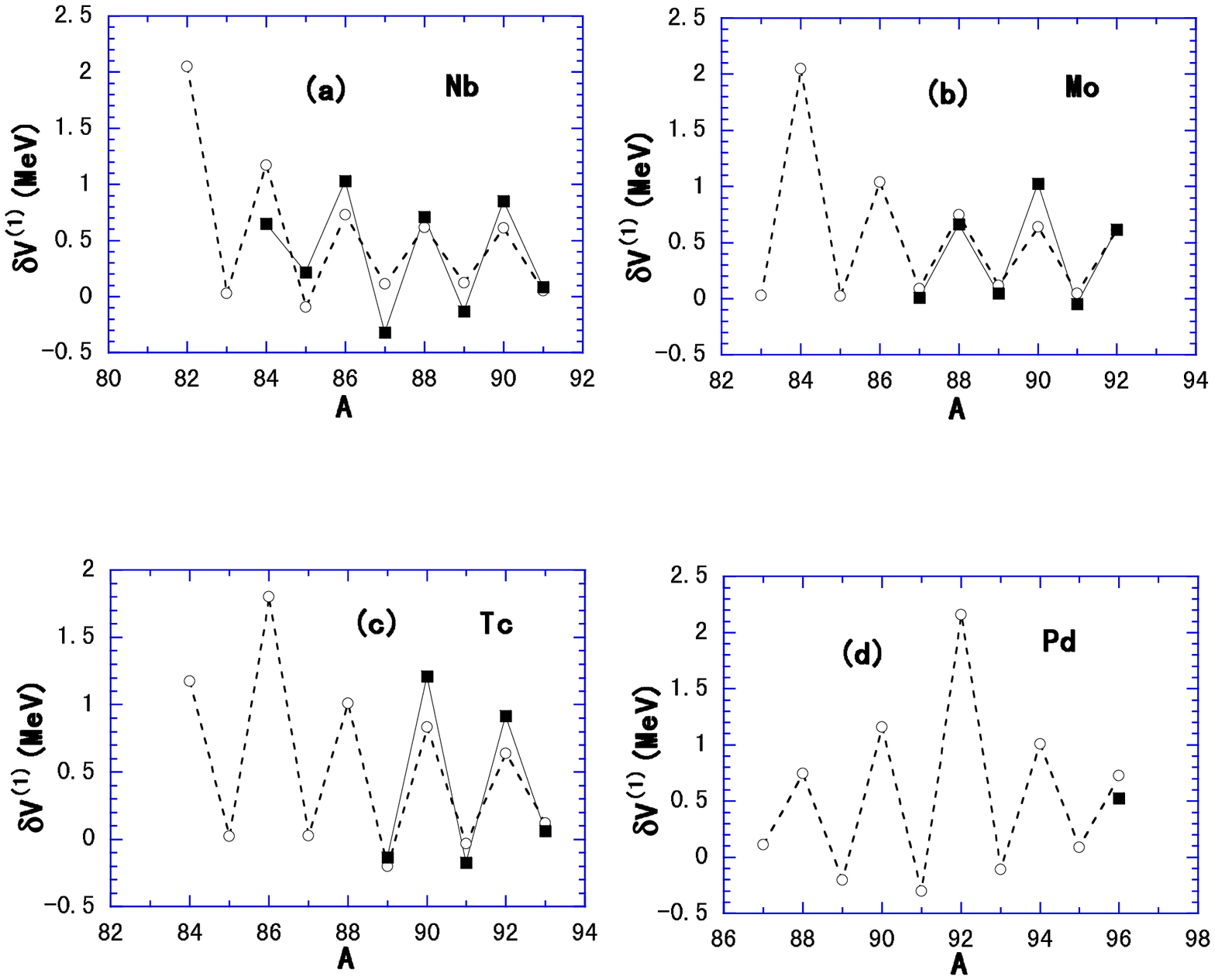}
\end{center}
\begin{center}
     \epsfxsize=17cm
     \epsfysize=22cm
     \epsfbox{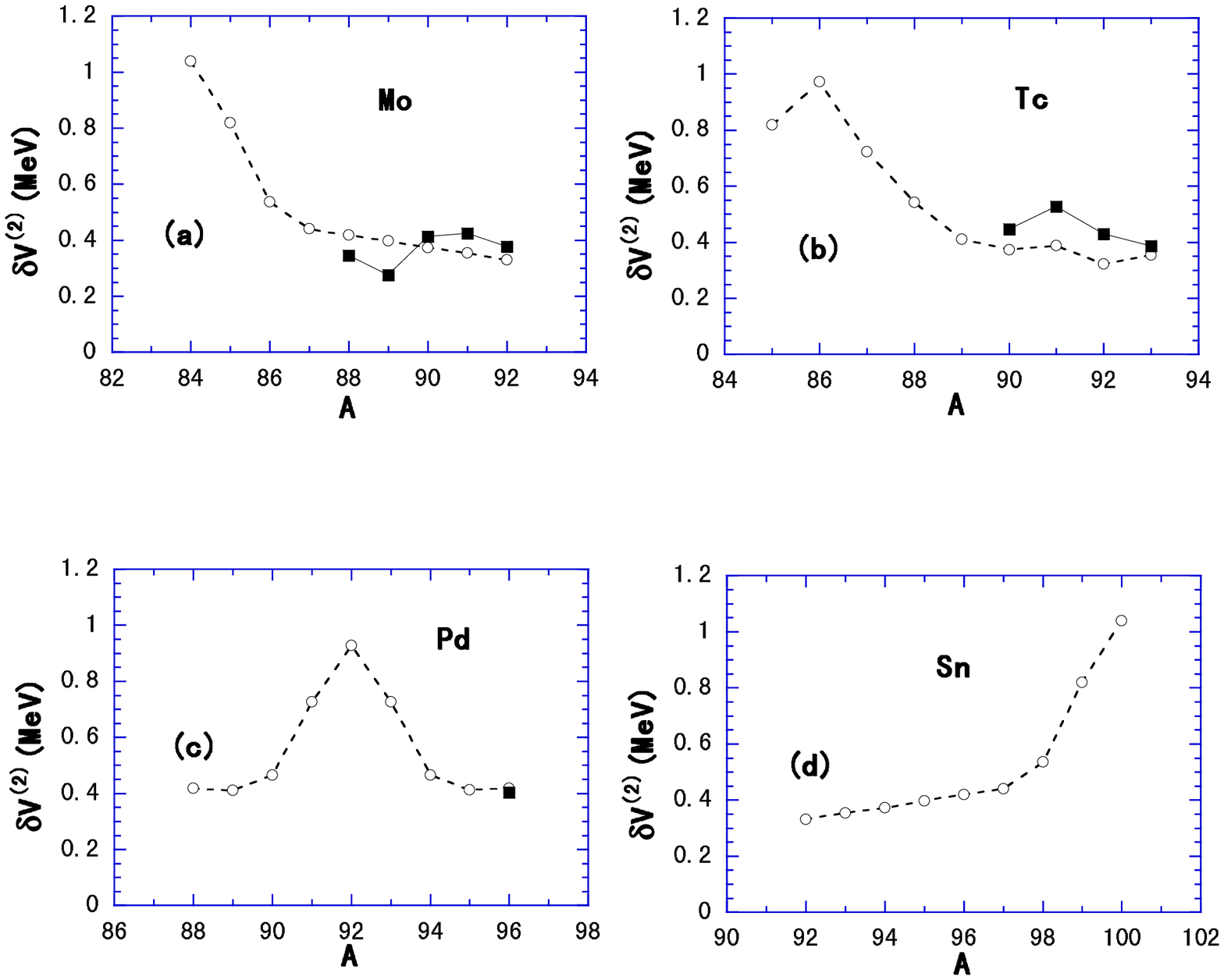}
\end{center}
\begin{center}
     \epsfxsize=17cm
     \epsfysize=22cm
     \epsfbox{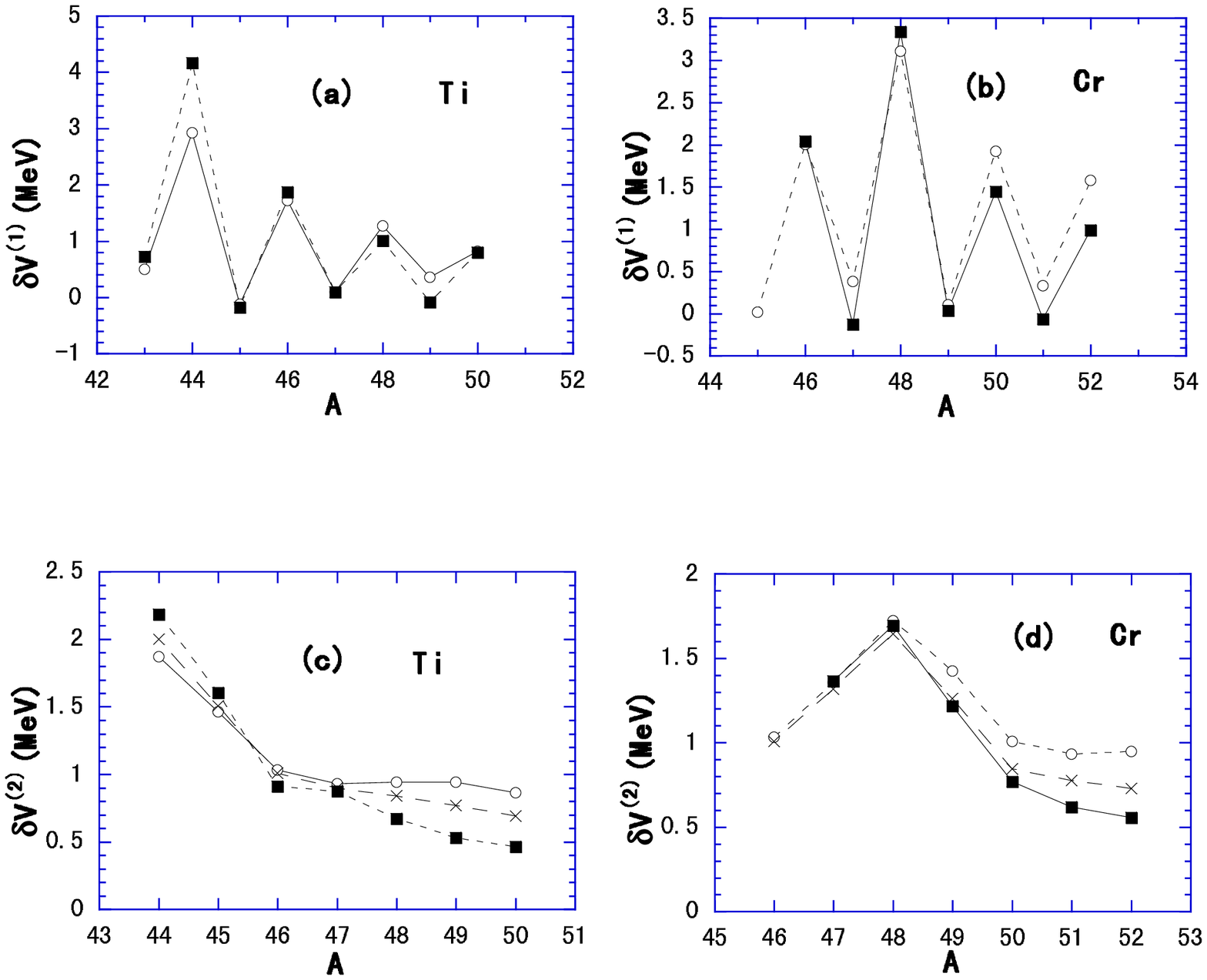}
\end{center}
\begin{center}
     \epsfxsize=17cm
     \epsfysize=20cm
     \epsfbox{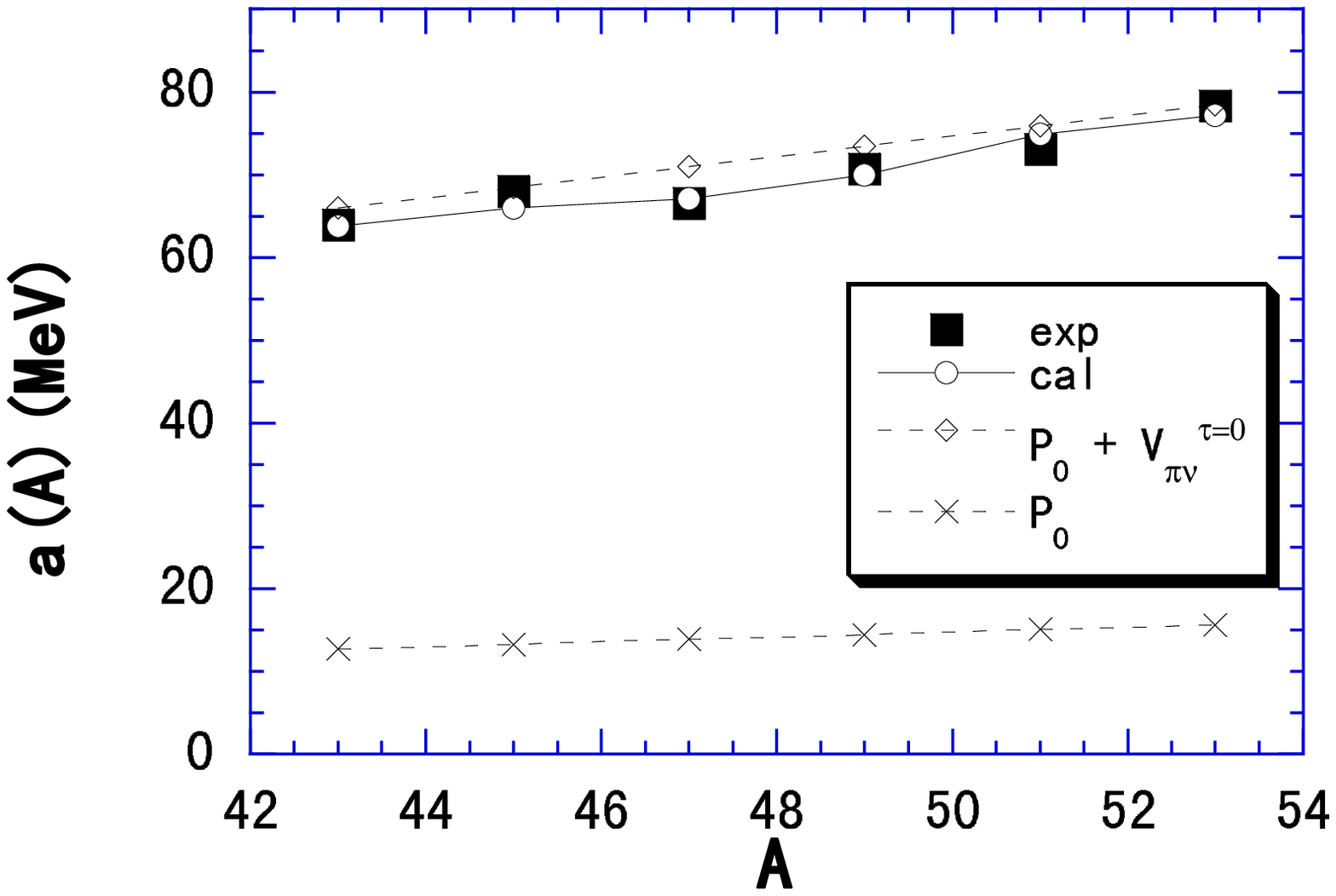}
\end{center}
\begin{center}
     \epsfxsize=17cm
     \epsfysize=17cm
     \epsfbox{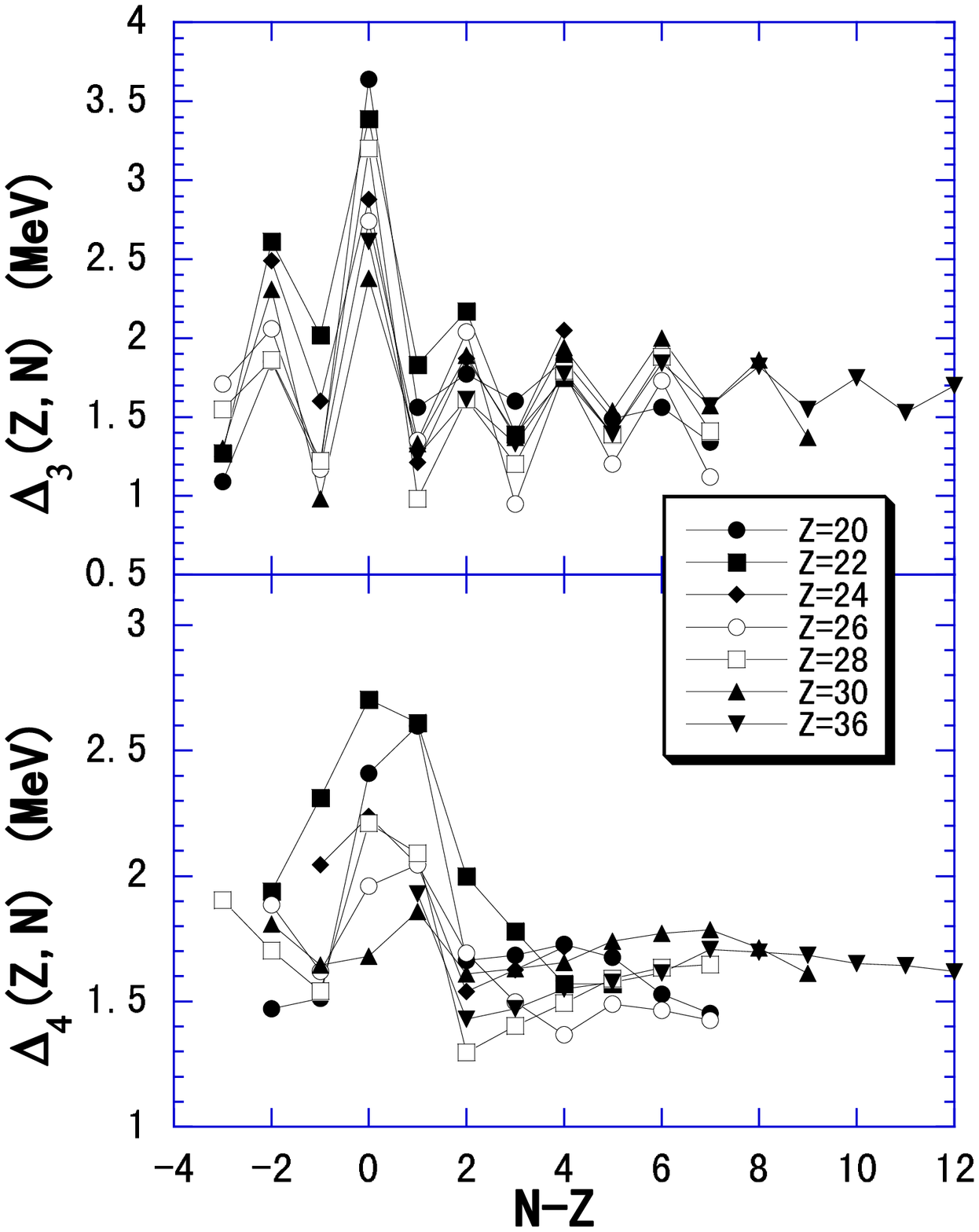}
\end{center}
\begin{center}
     \epsfxsize=17cm
     \epsfysize=17cm
     \epsfbox{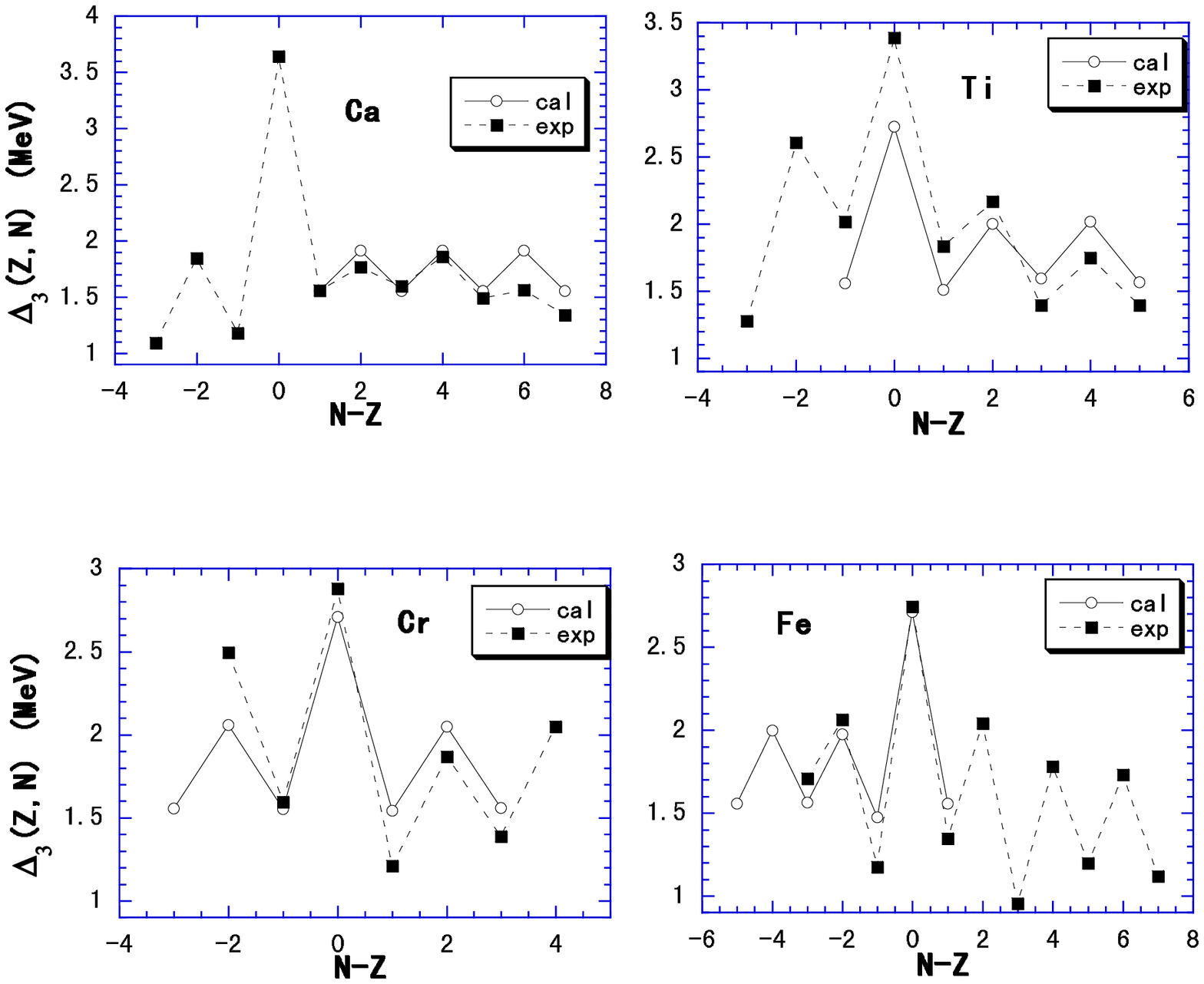}
\end{center}
\begin{center}
     \epsfxsize=17cm
     \epsfysize=17cm
     \epsfbox{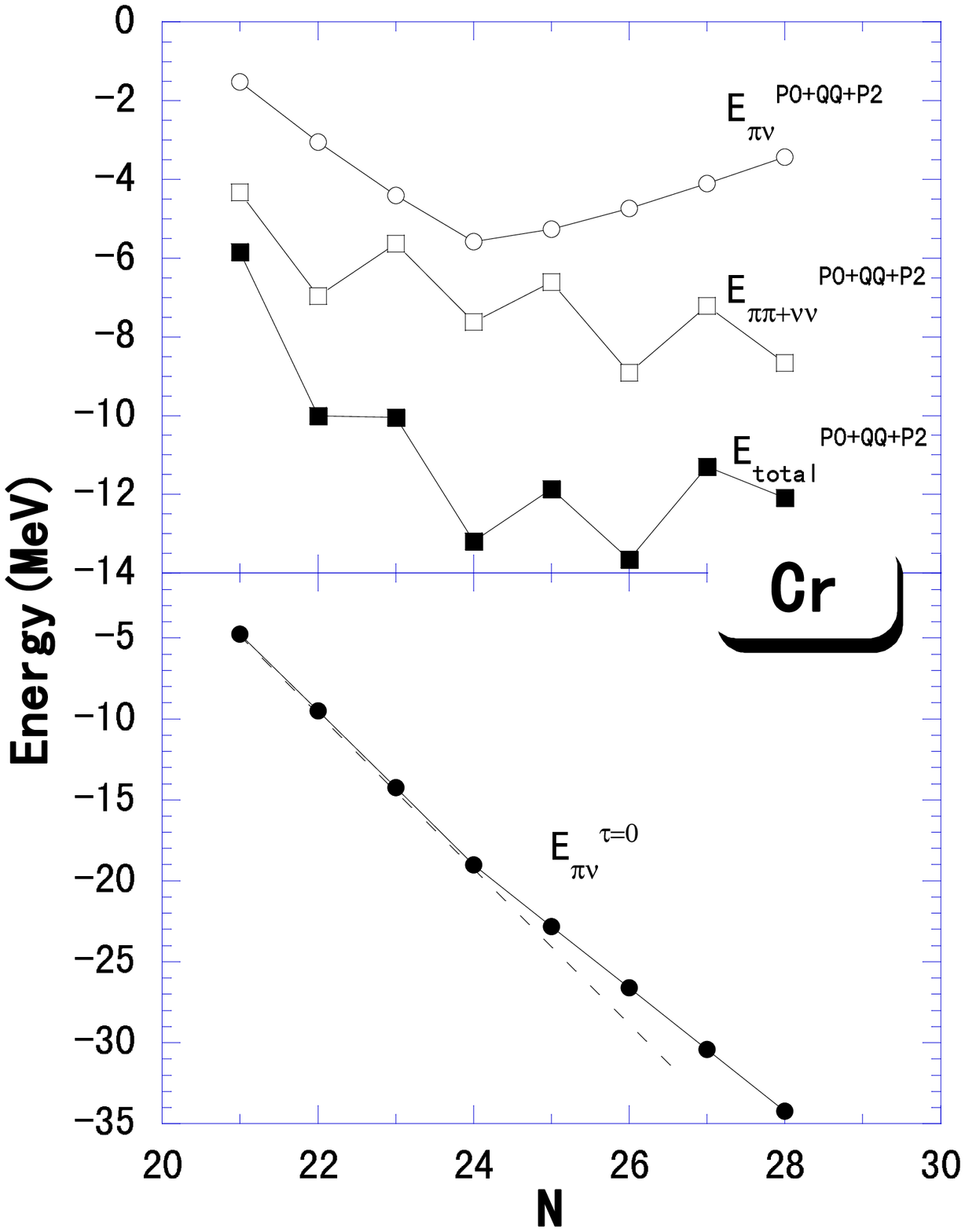}
\end{center}
\begin{center}
     \epsfxsize=17cm
     \epsfysize=17cm
     \epsfbox{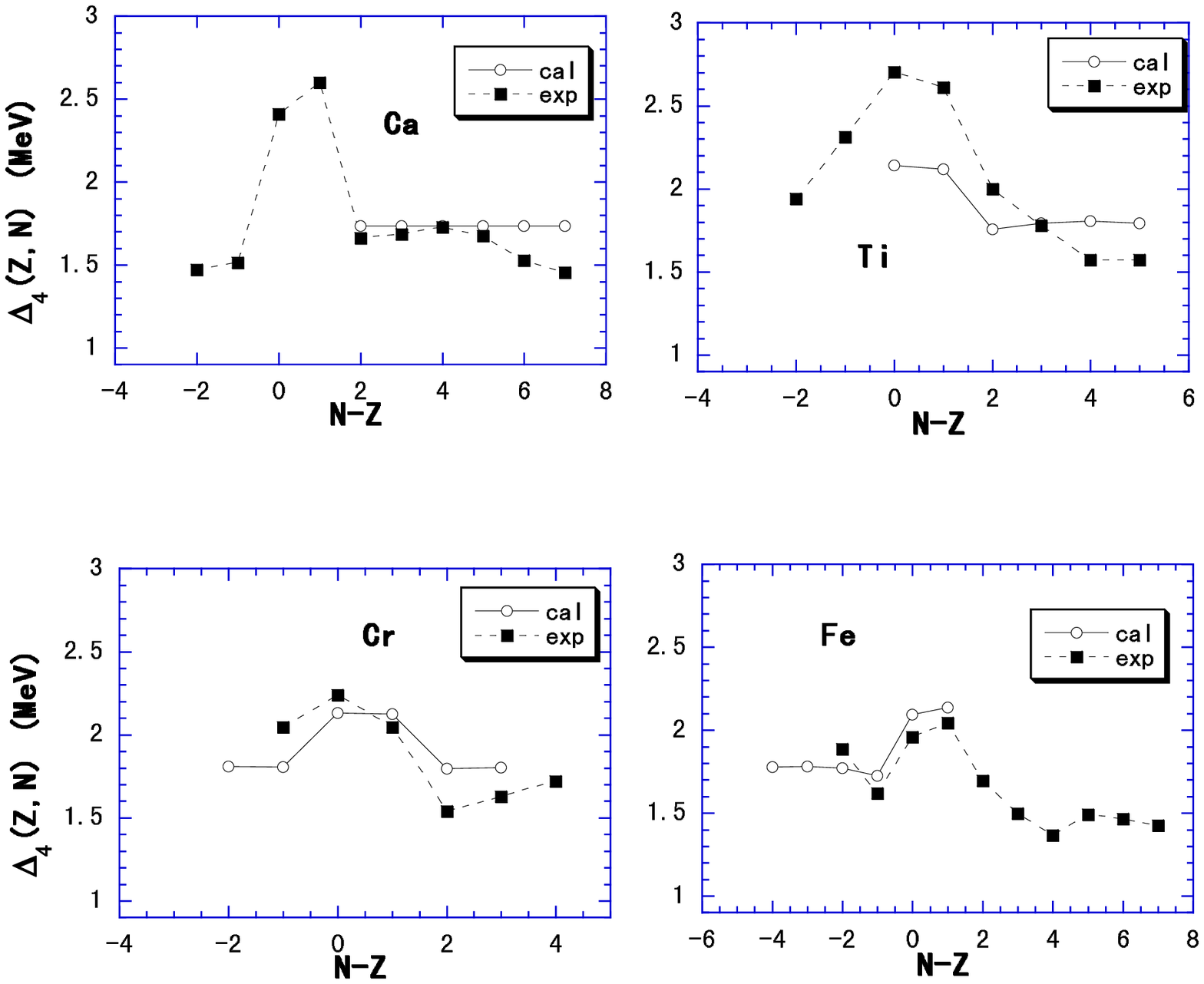}
\end{center}
\begin{center}
     \epsfxsize=17cm
     \epsfysize=25cm
     \epsfbox{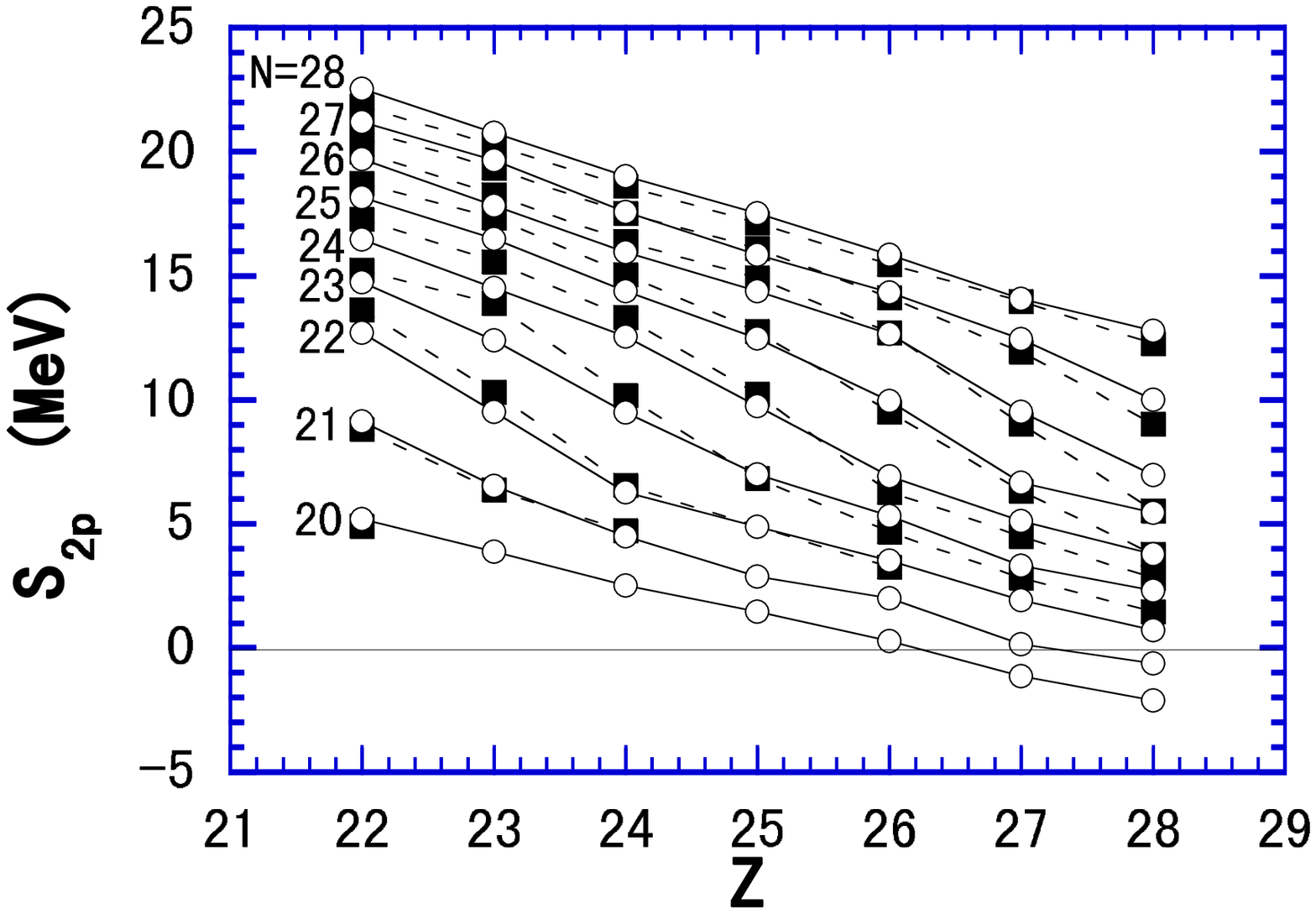}
\end{center}
\end{document}